# An Artificial Neural Network Based Approach for Identification of Native Protein Structures using an Extended ForceField

Timothy Matthew Fawcett[1], Stephanie Irausquin[1], Mikhail Simin[1], and Homayoun Valafar[1]
[1]Computer Science and Engineering, University of South Carolina, Columbia, SC, USA
fawcettt@cec.sc.edu, irausquin@biol.sc.edu, siminm@cec.sc.edu, and homayoun@cec.sc.edu

**Abstract**

*Current protein forcefields like the ones seen in CHARMM or Xplor-NIH have many terms that include bonded and non-bonded terms. Yet the forcefields do not take into account the use of hydrogen bonds which are important for secondary structure creation and stabilization of proteins. SCOPE is an open-source program that generates proteins from rotamer space. It then creates a forcefield that uses only non-bonded and hydrogen bond energy terms to create a profile for a given protein. The profiles can then be used in an artificial neural network to create a linear model which is funneled to the true protein conformation.*

## 1. Introduction

Proteins are essential biochemical compounds performing many vital processes within the cell. Yet the health and survival of all organisms is dependent on the ability of proteins to fold correctly into their native 3-dimensional structure so that they may carry out their particular function [1]. A number of diseases such as Alzheimer's, Parkinson's, Type II Diabetes, cystic fibrosis, and certain cancers are associated with mis-folded proteins [2-5]. Therefore furthering our understanding of the mechanisms involved in protein structure prediction and protein folding has never been so important.

The question of how a protein folds into its 3-dimensional atomic structure given its amino acid sequence is often referred to as the "protein folding problem" [6] and the concept of an energy landscape is fundamental to the mechanism of protein folding [7]. A protein contains a conformational state that can be linked to thermodynamics, with a protein obtaining its conformational fold when its energy is in it lowest state [8]. Protein folding from energetics alone is important for several reasons. Firstly, there are over 10,000 distinct fold families for proteins [9]. The Protein Data Bank (PDB) [10], however, contains approximately 1,500 fold families as reported by CATH [11] or SCOP [12]. Therefore structure prediction of proteins based on homology modeling will have a limited rate of success. Whereas the successful folding of proteins based purely on known physical forces and computational modeling will have far less limitations.

In addition, energy-based folding of proteins will allow for a better comprehension of the internal dynamics of proteins. Because proteins are not suspended in a specific state or in a specific medium, the energy forcefield can lead to a more accurate structure. This promotes better modeling and molecular dynamic simulations for proteins, especially those which exist in different conformations as exemplified by the human TS protein [13].

Furthermore, physics-based folding of proteins will prove useful in the realm of protein/protein and protein/ligand interactions, further contributing to our understanding of diseases at the molecular level. For example, sickle cell anemia is caused by a substitution of valine for glutamic acid at the sixth amino acid position in the β-globin gene [14], [15]. Whereas this single point mutation would have been inconsequential and unnoticed by homologous modeling approaches, the proper understanding of internal physical forces is more likely to identify the catastrophic effects of the single amino acid change. This may ultimately lead to effective forms of treatment and/or therapies.

Here we present a program which simplifies the forcefield, resulting in a smoother and more manageable energy landscape. This is accomplished through a reduction of the energy equation from a rotamer space. The use of the hydrogen bond term, as well as non-bonded energy functions, are then used to describe the protein and its energy landscape which funnel an amino acid sequence to a conformation by way of an artificial neural network.

## 2. Theoretical Background

### 2.1. Existing Models and Limitations

CHARMM and Xplor-NIH [16-18] use a similar equation for the protein forcefield (Equation 1). These terms should describe the protein completely and be able to find the protein's conformation, however this is not always the case for a number of reasons. First, the number of variables in the energy landscape are too vast and finding the global minimum is not possible with the minimization techniques that are currently available [19]. Secondly, the energy forcefield is often not understood well enough.

$$U(R) = \sum E_{bonds} + \sum E_{angles} + \sum E_{dihedrals} + \sum E_{impropers} + \sum E_{vdw} + \sum E_{elec} \quad (1)$$

Comparison of an experimentally obtained Ramanchandran plot versus its computational counter part can provide the means for evaluating our modeling of short range steric collisions. Current models of steric collision rely on a Lennord-Jones 12-6 representation of the Van der Waals interaction. Figures 2 and 3 illustrate experimentally and computationally computed Ramachandran maps respectively. The observable similarity between these two figures can serve as a confirmation that our modeling of steric interactions is relatively accurate. However, such comparisons can not be extended to all other individual force terms that describe protein or peptide geometries. There are many terms in equation 1 where constant values and measured values raise questions about our current understanding of the forcefield. In addition to accurate representation of physical forces, proper integration of force terms into one effective force potential is yet another standing challenge in purely-modeling protein structure determination. In summary, there are at least two issues which can be cited as impediments to computational modeling of protein structures:

1. Inadequacies in global optimization techniques that fall victim to the complexity of energy landscapes of proteins. This is especially problematic given the high-dimensionality of the search space.
2. Inadequate or incomplete representation of the forces that lead to an ineffective folding of proteins. It is clear that hydrophobicity/ hydrophilicity of amino acids, cooperative H-bonding, and other interactions with water molecules can play an important role in the folding of proteins. It therefore stands to reason that a more complete representation of these forces may reduce the complexity of the energy landscapes.

## 3. Methods
### 3.1 SCOPE

SCOPE (Semi Classical Open-source Protein Energy) is an open-source software program implemented in C++, which will be available for download in August 2011 at http://ifestos.cse.sc.edu. General implementation details have been presented previously [20]. One unique feature of SCOPE is its representation of protein structures in their rotamer space instead of the traditional Cartesian representation. SCOPE is also capable of calculating non-bonded forces from the reconstructed protein structure. Reconstruction of the protein from the rotamer space is important because this allows for a reduction in the number of parameters used to create and calculate the energies of a protein. A reduction in dimensionality of the search space helps to mitigate the first noted impediment in Section 2.1. For example, to model a single Lysine amino acid, Cartesian representation would require the x, y, and z coordinates for each atom. Since Lysine contains 22 atoms, this produces 66 (22*3) parameters to construct. In contrast, the rotamer representation of the same amino acid space would require a total of 7 parameters: the $\phi$, $\psi$, $\omega$, and the four $\chi$ angles. Therefore, SCOPE boasts a significant reduction in the number of parameters needed to reconstruct an entire protein. In addition, use of the rotamer space preserves the perfect peptide geometries, therefore eliminating the need to calculate bonded energies such as: $E_{bonds}$, $E_{angles}$, $E_{impropers}$ [20]. Furthermore, these calculated non-bonded energies are comparable to those obtained with CHARMM [20]. More recently, we have included the addition of a hydrogen bond term in SCOPE to aid with protein refinement.

### 3.1.1 Extended Forcefield by Inclusion of an Explicit H-bond Term

In contrast to covalent bonds, a hydrogen bond consists of electrons which are shared between atoms with compatible electron affinities [21], [22]. In a hydrogen bond the hydrogen atom is typically referred to as the donor, while a second and more electronegative atom (such as an Oxygen or Nitrogen) is denoted as the acceptor. Inclusion of a third atom, to which the hydrogen is covalently bonded, forms the three atoms that define a hydrogen bond.

The formation of hydrogen-bonds in α-helices and β-sheets was first speculated by Linus Pauling before being discovered by X-ray crystallography [23]. Hydrogen bonds are very critical in formation and stabilization of secondary structures of proteins. They are also among the strongest non-bonded interactions [22]. Therefore their explicit representation in an extended forcefield is easily justified.

The hydrogen bond energy term has not been explicitly included in traditional forcefields. This exclusion is based on the argument that a combination of electrostatic and Van der Waals interactions should implicitly encapsulate the hydrogen bond term. However, since hydrogen bonds consist of both distance and orientational components, it can be speculated that hydrogen bond energies are not properly evaluated. It is possible for a hydrogen bond not to be formed as a result of failing to satisfy the orientational requirements, despite satisfying the distance requirements [22]. Furthermore, consecutive hydrogen bonds have been shown to exhibit a cooperativity phenomenon, where the total potential energy is greater than the sum of its individual components [22].

### 3.1.3. Scope Hydrogen Bonding

Explicit calculation of hydrogen bonds is now available within SCOPE and our implementation is designed to be consistent with that of the DSSP (Define Secondary Structure Prediction) program [24]. Hydrogen bond energy is calculated using a formula (Equation 2) that varies the distance and angular components to allow for errors in measurements of the atoms. The variables $e_1 = 0.42$ and $e_2 = 0.20$ are charges on the atoms, *r* is the distance between particular atoms, and *f* is a dimensional factor that is set to a constant value of 332 [24]. Bifurcation of hydrogen bonds is permitted, in which case the energy for two hydrogen bonds per residue are computed.

$$E_{HBOND} = e_i e_j f * (\frac{1}{r(ON)} + \frac{1}{r(CH)} - \frac{1}{r(OH)} - \frac{1}{r(CN)}) \quad (2)$$

SCOPE only calculates hydrogen bonds for the backbone atoms. The values are calculated for each amino acid against every other amino acid and stored in a table. The table is then searched row by row to find the lowest two values. There are two exceptions for which a value may not be considered to be one of the lowest values. The first exception is that a hydrogen bond cannot exist on the same amino acid; for example, an amide group cannot hydrogen bond with the carboxyl group of the same amino acid. The second exception is that it is not possible to have a hydrogen bond between an adjacent amino acid [24]. A hydrogen bond only exists if the energy calculated is less than or equal to -0.5 kilocalories per mole [24]. Not only does SCOPE calculate the hydrogen bond energy values, but it also reports the total number of hydrogen bonds in the protein. Secondary structures are built by consecutive hydrogen bonds in amino acids. SCOPE breaks down the hydrogen bonds into sections where there are consecutive hydrogen bonds in the protein thereby separating the protein into sections of secondary structures (Figure 1).

### 3.1.4. SCOPE's Energy Profile

```
24         1    2    2    1    3    2    4    1    5
1          6    1    7    5    8    2    9    19   10
2          11   2    12   2    13   17   14   11   15
1          16   1    17   1    18   3    19   5    20
1          21   2    22   3    23   16   24   1
test0.ang  1.2861e+12  1.2861e+12   -7282.159682  20190.355307  102
```

Figure 1. An example of the SCOPE output.

SCOPE creates an energy profile for each protein input by the user. The profile begins with the total number of sections with consecutive hydrogen bonds followed by the number of consecutive bonds in each section (Figure 1). The file name and the energy for the protein is given on the next line where the energies include: the total, Van der Waals, electrostatic, and hydrogen bond energies and is followed by the total number of hydrogen bonds in the protein (Figure 1). The hydrogen bond profile is determined based on the number of ungapped and consecutive hydrogen bonds that are formed along the backbone of the protein. The least number of sections per protein could be zero (no hydrogen bonding), while the maximum number of sections may be n/2 where n is the number of amino acids in the protein.

### 3.2 Artificial Neural Network

Conventional approaches for evaluation of protein structures has been based on calculating a total energy term for the entire protein. Although simple, this approach may lack the ability to decipher severe internal problems. Here we have utilized an Artificial Neural Network (ANN) approach to evaluate structural fitness. More specifically, the SCOPE profiles have been used as an input to ANN for classification of its viability as the native structure. An input pattern includes the total number of hydrogen bonds, the number of sections with consecutive hydrogen bonds, the consecutive hydrogen bonds in each section, a scaled Van der Waals energy, and a scaled electrostatic energy. A scaled backbone root mean squared deviation (BB RMSD) is used as the desired output. The Van der Waals and electrostatic energies are scaled to a value between 0 and 10 inclusive (Equations 3 and 4). This scaling is implemented so that the lower the energy, the higher the scaled score. The BB RMSD is calculated against the true structure using MolMol [25] and is scaled to a value from 0 to 1 by dividing each one by the maximum BB RMSD of the complete set of proteins used in the experiment.

$$Scaled\ VDW = (\frac{\log(Max) - \log(VDW)}{\log(Max) - \log(Min)}) * 10 \quad (3)$$

$$Scaled\ Elec = (\frac{Max\ Elec - Elec}{Max\ Elec - Min\ Elec}) * 10 \quad (4)$$

The artificial neural network architecture that was used here consisted of a feed forward network (FFN) with one hidden layer. The number of inputs will depend on which protein is being examined through the network. The number of inputs will be the total number of hydrogen bonds, the number of sections, each section's consecutive hydrogen bonds, the scaled Van der Waals term, and the scaled electrostatic term. The nodes on the input layer and hidden layer use a tansig function (Equation 5). The optimization method is the Levenberg-Marquardt algorithm [26]. There is only one output node that will represent the scaled BB RMSD. The output node has a pure linear function instead of the tansig function.

$$a = \frac{e^n - e^{-n}}{e^n + e^{-n}} \quad (5)$$

### 3.3 Evaluation and Testing Procedures
### 3.3.1 Ramachandran Plot with SCOPE

The limitations in the space of amino acid torsion angles was noted as early as 1963 by G. N. Ramachandran [22]. This observation has resulted in a Ramachandran plot that illustrates the region of acceptable ϕ and ψ dihedral angles of amino acids (Figure 2). Our first exercise in validating implementation of physical forces within SCOPE is to reproduce this pattern computationally. This is primarily to test of the non-bonded energy terms. A tri-alanine was used for this exercise, where the first and third residues were fixed with their φ set to -60 and their ψ set to 120. The middle alanine's φ and ψ were rotated from -180 to 180 using 1 degree increments and the Van der Waals energy was recorded. All the data points were plotted and a 3-D heat map was created using gnu plot with an applied threshold of 2 kilocalories.

### 3.3.2. Energy Based Identification Of The Most Viable Structure

To study molecular energy as a function of backbone RMSD to the reference structure, one thousand perturbations of 1A1Z were created. These structures, along with the original structure varied between 0 – 4.47Å, and were subjected to energy evaluation by SCOPE. The total energy for each of the structures was recorded along with the BB RMSD to the true structure. A graph of the correlation between the BB RMSD and energies was constructed. An effective mechanism of identifying the most viable structure should produce a funneling effect leading to the structure with the least degree of structural difference to the actual structure.

### 3.3.3. Testing Based on ANN Classification

The Neural Network was tested on 3 proteins with different secondary structure characteristics, 1A1Z (the same protein as used in section 3.3.2), 2PTV, and 1G10. Where 1A1Z is an α-helical structure, 2PTV is a β-sheet structure, and 1G10 is a mix of α-β structure. Each of the proteins had 1000 perturbed structures that were generated by rotating their φ and ψ dihedral angles. The proteins BB RMSD varied between 0 – 5.5Å inclusive to the original structure. A total of 1001 structures (the 1000 perturbed and the original structure) were used to form the data for testing and training of the neural network.

All 1001 proteins were run through SCOPE in order to create an energy profile for each structure. The energy profiles were all placed into a .csv file. Once in the .csv file the Van der Waals, electrostatic, and BB RMSD values were scaled. Finally, the energy profiles were split into three sets for the neural network. The training set consisted of the first 500 records (in a randomized list). The second set consisted of 167 records that were used as the validation set. The test set consisted of the final 334 records as well as the true structure.

The neural network used is a part of the gnu octave program. Each protein was tested until 10 results ended in a validation stop. The mean squared error (MSE) of the neural network was recorded and examined after each run to ensure consistent outcomes. The results of the neural network were then compared to the actual BB RMSD of each structure in the testing set. A linear correlation was observed (Figure 6) for the testing set and the numerical rank of the correct structure was also recorded.

## 4. Results
### 4.1. Ramachandran Plot

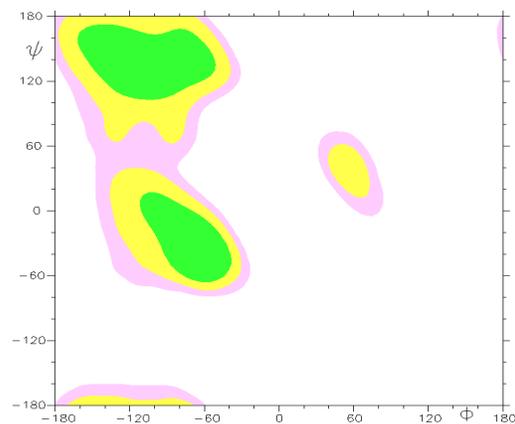

Figure 2. Ramachandran plot of the amino acid alanine created in MolMol.

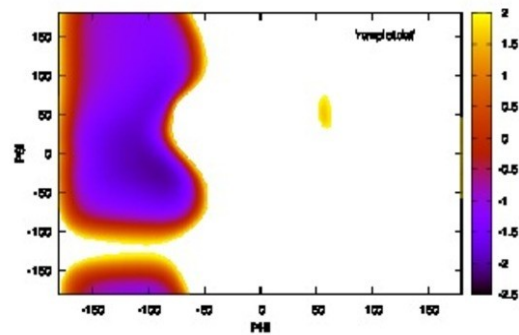

Figure 3. Ramachandran plot of the tri-alanine amino acid sequence energies calculated using SCOPE.

Figure 2 represents the experimental Ramachandran plot that is included as part of MolMol [25] using the alanine amino acid. Figure 3 represents the computational Ramachandran plot that was created in SCOPE. There is a significant degree of similarity between the two maps. Any differences can be attributed either to evolutionary selection mechanisms or steric contribution due to larger proteins.

### 4.2. Energy Based Evaluation of Structural Conformers

Figure 4 shows the results of 1000 derivative structures of 1A1Z. Each protein was evaluated with SCOPE and their total energies were recorded. In addition, the BB-rmsd of each structure was calculated with respect to the actual 1A1Z structure. The resulting graph was studied in search of a funneling effect for the known energy forcefield. This figure exhibits a lack of the desired funneling effect. In fact, some of the structures that are far from the original structure exhibit the lowest energies.

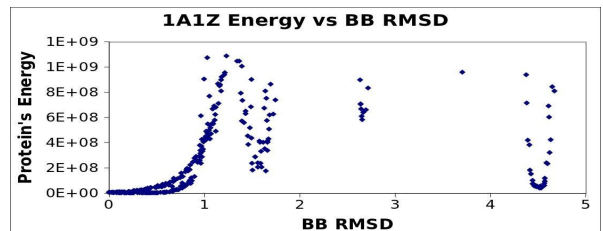

Figure 4. The BB RMSD vs the energy of 1000 different structures of 1A1Z.

### 4.3. Neural Network Results

The first test for the neural network was conducted using the 1A1Z protein (Figure 5). There were 334 protein structures in this testing set varying within 0 – 4.47 Å from the true structure. The number of inputs for the neural network was 14. The true structure was given a score of 0.139599673 from the neural network and was ranked first with 64 other structures.

The second test on the neural network was run on protein 2PTV (Figure 6). For this particular protein there were 334 proteins in the test set varying within 0 – 5.044 Å from the true structure. There were 23 inputs for the neural network and the true structure was picked fourth overall with a score of 0.019404271.

The third test for the neural network was run using protein 1G10 (Figure 7). Here there were 334 proteins in the test set varying from within 0 – 5.496 Å of the true structure. The number of outputs for the neural network was 30. The true structure was picked to be 16 overall and had a value of 0.097540688.

### 5. Discussion

The agreement between the Ramanchandran plot produced by SCOPE and that of the experimental data, confirms SCOPE's representation of Van der Waals and electrostatic forces. Despite the accuracy in representation of the non-bonded forces and ideal representation of peptide geometries, we have demonstrated the lack of any funneling effect when using traditional forces for identification of the native structure of proteins. In theory, it is reasonable to argue that if two structures have a BB RMSD that is less than 1Å then the two structures should have similar energies. Also if two structures are far apart, with a BB RMSD of 4Å for example, then the energies should be distinguishably different. Figure 4 shows the BB RMSD vs. the calculated energy of the protein 1A1Z. Here there are structures that are approximately 4.5Å from the true structure and yet their energies are close to the native conformation of this protein. This illustrates that the known forces in the forcefield equation are not enough to successfully guide a protein to its native conformation. Therefore the traditional representation of physical forces are inadequate in guiding de novo protein folding.

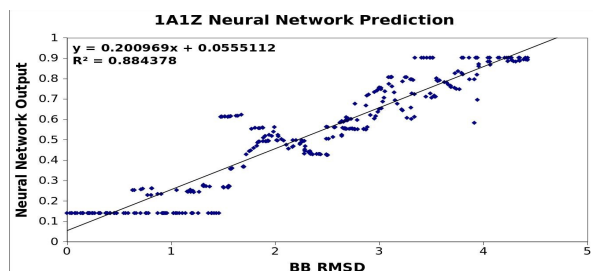

Figure 5. The results of the neural network for protein 1A1Z. The true structure was tied for first place with 64 other structures.

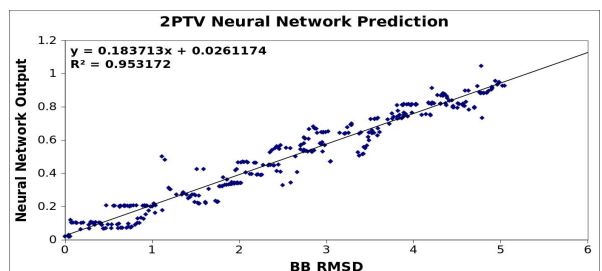

Figure 6. The neural network output for protein 2PTV. The true structure was ranked 4th overall.

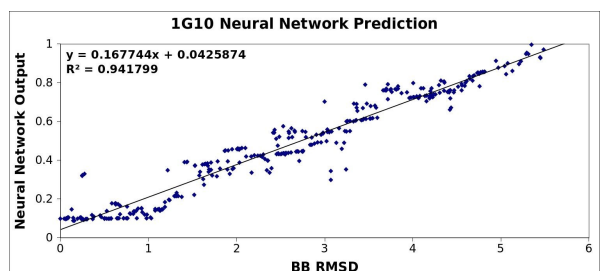

Figure 7. The neural network output for the protein 1G10. The true structure was ranked 16th overall.

Our artificial neural network based approach combined with representation, in addition to the explicit representation of the hydrogen-bond energy term, exhibits a much better funneling effect in guiding the folding or refinement of protein structures. Results from the 1A1Z protein demonstrate a clustering of the protein and a remarkable linear funneling of the structures, with the true structure of 1A1Z being tied for first place with 64 other competitive structures. A close examination of the top 65 structures shows that the structures varied from 0 to 1.46Å. Therefore, the addition of the explicit hydrogen bond information is instrumental for the neural network to identify structures that are very close to the original structure.

The output from the neural network for protein 2PTV similarly shows a clustering of the proteins in a linear model. The true structure was ranked fourth overall, with the remaining three structures exhibiting a BB RMSD between 0.039 and 0.049Å from the original structure – a difference so small that these four structures can be considered identical.

Likewise, the last experiment from the neural network conducted with protein 1G10, revealed a linear clustering of the structures. The fifteen structures that were picked above the true structure varied from 0.044 to 0.366Å. Once again these fifteen other structures are so close that they can be considered to be the same as the true structure.

The neural network shows very promising results. The addition of the consecutive hydrogen bonds has helped to determine the correct structure for a protein. The hydrogen bond is modeled after the DSSP program and is accurate in most cases. Although DSSP determines when steric collisions in the backbone atoms are present, this is not considered in SCOPE; this is due to the fact that in such cases the Van der Waals calculation is extremely high, thereby determining that the current structure is not correct.

Although the neural network approach has demonstrated significant success in predicting the fitness of a structure, additional tests are needed to determine the sensitivity of the neural network. Such tests will be conducted to demonstrate the artificial neural network's ability in generalization, thereby excluding any possibility of memorization. To that end, we will test the performance of our artificial neural network based approach on proteins that were not included during the training phase. As part of our future work we plan to add hydrogen bonding of the side-chain atoms and a hydrophobicity term.

## 6. Acknowledgements

We would like to thank Dalton Brown for helping with source code development. This work was supported by grant No. NIH-1R01GM081793 and the SC INBRE grant No. P20 RR-016461.